\documentstyle[sprocl]{article}

\def\ci#1{\cite{#1}}
\def\bi#1{\bibitem{#1}}
\def\vak{V.A.\ Kosteleck\'y}
\def\al{\alpha}
\def\be{\beta}
\def\ga{\gamma}
\def\de{\delta}

\def\la{\lambda}

\def\si{\sigma}

\def\ps{\psi}
\def\om{\omega}

\def\De{\Delta}

\def\cE{{\cal E}}
\def\ea{{\it et al.}}
\def\fr#1#2{{{#1} \over {#2}}}
\def\ap{\al^\prime}

\def\ket#1{|{#1}\rangle}

\def\half{{\textstyle{1\over 2}}}
\def\frac#1#2{{\textstyle{{#1}\over {#2}}}}
\def\lsim{\mathrel{\rlap{\lower4pt\hbox{\hskip1pt$\sim$}}
    \raise1pt\hbox{$<$}}}
\def\gsim{\mathrel{\rlap{\lower4pt\hbox{\hskip1pt$\sim$}}
    \raise1pt\hbox{$>$}}}
\def\hydrogen{H}
\def\antihydrogen{$\overline{\rm{H}}$}
\def\h{\hydrogen}
\def\ah{\antihydrogen}
\newcommand{\beq}{\begin{equation}}
\newcommand{\eeq}{\end{equation}}
\newcommand{\bea}{\begin{eqnarray}}
\newcommand{\eea}{\end{eqnarray}}
\newcommand{\rf}[1]{(\ref{#1})}

\def\ajm #1 #2 #3 {Am.\ J.\ Math. {\bf #1}, #3 (18#2)} %
\def\ajp #1 #2 #3 {Am.\ J.\ Phys.\ {\bf #1}, #3 (19#2)}
\def\ant #1 #2 #3 {At. Dat. Nucl. Dat. Tables {\bf #1}, #3 (19#2)}
\def\ap #1 #2 #3 {Ann.\ Physics\ {\bf #1}, #3 (19#2)} %
\def\apb #1 #2 #3 {App. Phys. B {\bf #1}, #3, (19#2)}    %
\def\app #1 #2 #3 {Appl.\ Physics\ {\bf #1}, #3 (19#2)} %
\def\baps #1 #2 #3 {Bull.\ Am.\ Phys.\ Soc.\ {\bf #1}, #3 (19#2)} %
\def\hfi #1 #2 #3 {Hyperfine Int.\ {\bf #1}, #3 (19#2)}
\def\ibid #1 #2 #3 {\it ibid., \rm {\bf #1}, #3 (19#2)}  %
\def\ijqc #1 #2 #3 {Internat.\ J.\ Quantum\ Chem.\
  {\bf #1}, #3 (19#2)} %
\def\jmp #1 #2 #3 {J.\ Math.\ Phys.\ {\bf #1}, #3 (19#2)}
\def\jms #1 #2 #3 {J.\ Mol.\ Spectr.\ {\bf #1}, #3 (19#2)} %
\def\jpa #1 #2 #3 {J.\ Phys.\ A {\bf #1}, #3 (19#2)}  %
\def\jpg #1 #2 #3 {J.\ Phys.\ G {\bf #1}, #3 (19#2)}  %
\def\jpsj #1 #2 #3 {J.\ Phys.\ Soc.\ Japan {\bf #1}, #3 (19#2)}
\def\lnc #1 #2 #3 {Lett.\ Nuov.\ Cim. {\bf #1}, #3 (19#2)}
\def\mj #1 #2 #3 {Math. Japon. {\bf #1}, #3 (19#2)} %
\def\mpl #1 #2 #3 {Mod.\ Phys.\ Lett.\ A {\bf #1}, #3 (19#2)}
\def\nat #1 #2 #3 {Nature {\bf #1}, #3 (19#2)}
\def\nc #1 #2 #3 {Nuov.\ Cim.\ A{\bf #1}, #3 (19#2)}
\def\ncb #1 #2 #3 {Nuov.\ Cim.\ B{\bf #1}, #3 (19#2)}
\def\nim #1 #2 #3 {Nucl.\ Instr.\ Meth.\ B{\bf #1}, #3 (19#2)}
\def\nima #1 #2 #3 {Nucl.\ Instr.\ Meth.\ Phys.\ Res.\
     A{\bf #1}, #3 (19#2)}
\def\npb #1 #2 #3 {Nucl.\ Phys.\ B{\bf #1}, #3 (19#2)}
\def\pha #1 #2 #3 {Physica \ {\bf #1}, #3 (19#2)}
\def\pjm #1 #2 #3 {Pacific J.\ Math.\ {\bf #1}, #3 (19#2)} %
\def\pla #1 #2 #3 {Phys.\ Lett.\ A {\bf #1}, #3 (19#2)}
\def\plb #1 #2 #3 {Phys.\ Lett.\ B {\bf #1}, #3 (19#2)}
\def\pp #1 #2 #3 {Phys.\ Plasmas\ {\bf #1}, #3 (19#2)} %
\def\prev #1 #2 #3 {Phys.\ Rev.\ {\bf #1}, #3 (19#2)}  %
\def\prep #1 #2 #3 {Phys.\ Rep. {\bf #1}, #3 (19#2)}  %
\def\pra #1 #2 #3 {Phys.\ Rev.\ A {\bf #1}, #3 (19#2)}  %
\def\prd #1 #2 #3 {Phys.\ Rev.\ D {\bf #1}, #3 (19#2)}
\def\prl #1 #2 #3 {Phys.\ Rev.\ Lett.\ {\bf #1}, #3 (19#2)}
\def\prs #1 #2 #3 {Proc.\ Roy.\ Soc.\ (Lon.) A {\bf #1}, #3 (19#2)}
\def\ptp #1 #2 #3 {Prog.\ Theor.\ Phys.\ {\bf #1}, #3 (19#2)}
\def\rmp #1 #2 #3 {Rev.\ Mod.\ Phys.\ {\bf #1}, #3 (19#2)} %
\def\sciam #1 #2 #3 {Sci.\ Am. {\bf #1}, #3 (19#2)} %
\def\zap #1 #2 #3 {Z.\ Angew.\ Phys.\ {\bf #1}, #3 (19#2)} %
\def\zn #1 #2 #3 {Z.\ Naturforsch.\ {\bf #1}, #3 (19#2)} %
\def\zp #1 #2 #3 {Z.\ Physik.\ {\bf #1}, #3 (19#2)} %

\def\al{\alpha}

\def\be{\begin{equation}}
\def\ee{\end{equation}}
\def\bea{\begin{eqnarray}}
\def\eea{\end{eqnarray}}

\begin{document}

\title{Bounding CPT- and Lorentz-Violating Parameters in a Standard-Model
Extension\footnote{Invited talk at the \it International Workshop on
the Lorentz Group, CPT, and Neutrinos\rm , Zacatecas, Mexico, June 1999.}}

\author{Neil Russell}

\address{Physics Department, Northern Michigan University,\\
Marquette, MI 49855, USA\\E-mail: nrussell@nmu.edu}

\maketitle

\abstracts{A general theoretical framework
that incorporates possible CPT and Lorentz violation
in an extension of the standard model and
in quantum electrodynamics has been developed
over the last decade.
The framework originates in the idea that
CPT and Lorentz symmetry could be broken spontaneously
in a more fundamental theory such as string theory.
These symmetry violations are described
in the standard-model extension
by small terms in the Lagrangian.
Various experiments can bound these quantities.
They include Penning-trap experiments
with electrons and positrons,
Penning-trap experiments with protons and antiprotons,
and possible experiments with hydrogen and antihydrogen
in traps or beams.
I will review aspects of the theory,
outline estimated bounds attainable
in specific experiments,
and present known bounds from completed experiments.}

\section{Introduction}

Numerous experiments have searched for violations of CPT
and Lorentz symmetry.
These include tests with
electrons and positrons \ci{vd}
and kaons.\ci{sch}
Up until recently,
no theoretical framework incorporating such violations
has been available to provide suggestions for
promising signals from the theoretical standpoint.
This proceedings
will provide an overview of work
done in collaboration with
Alan Kosteleck\'y and
Robert Bluhm
aimed at identifying interesting
experimental tests
in particle-trap systems
from theoretical considerations.
The context
is an extension of the standard model
of particle physics and
electrodynamics \ci{ck98}
that includes the possibility of
small violations of CPT and
Lorentz symmetry
while maintaining the important conventional features of
the standard model.
Several experiments searching for
the miniscule effects in this theoretical context
have been performed,
and are briefly mentioned.
The framework has recently been used to investigate
signals of Lorentz and CPT violation
in atomic clock-comparison experiments \ci{kla}
and the muon system.\ci{bkl99}
References on the standard-model extension
and a broader overview
of experimental tests may be found in
a recent review.\ci{k}

\section{Standard-Model Extension}

The origin for the extension of
the SU(3)$\times$SU(2)$\times$U(1) standard model
and quantum electrodynamics
is the concept of spontaneous CPT and Lorentz breaking
in a fundamental framework such as
string theory.\ci{kp1,ks}
This framework lies within the context of
conventional quantum field theory
and appears to preserve a number of
important conventional features
of the standard model such as gauge invariance,
power-counting renormalizability,
and microcausality.
Possible violations of CPT and Lorentz symmetry
enter into the theory through couplings
that could potentially give experimental signals.
Experiments with the potential to place bounds
on the couplings include investigations
with trapped hydrogen (\hydrogen)
or antihydrogen (\ah),
and with Penning traps
to be discussed here.

In this context,
the Dirac equation
obeyed by a four-component spinor field $\ps$
describing a particle with charge $q$ and mass $m$
has the form\ci{ck98,bkr98}
\bea
\left( i \ga^\mu D_\mu - m - a_\mu \ga^\mu
- b_\mu \ga_5 \ga^\mu - \half H_{\mu \nu} \si^{\mu \nu}
+ i c_{\mu \nu} \ga^\mu D^\nu
+ i d_{\mu \nu} \ga_5 \ga^\mu D^\nu
\right.
&&
\nonumber \\
\left.
+ i e_{\mu}D^\mu - f_\mu \ga_5 D^\mu
+ \half i g_{\mu\nu\la}\si^{\mu\nu} D^\la \right) \ps = 0
&& .
\label{dirac}
\eea
Here,
$i D_\mu \equiv i \partial_\mu - q A_\mu$
and $A^\mu$ is the electromagnetic potential.
The equation is given in terms of a set
of effective coupling constants,
$a_\mu$, $b_\mu$,
$H_{\mu \nu}$, $c_{\mu \nu}$, $d_{\mu \nu}$,
$e_\mu$, $f_\mu$, and $g_{\mu\nu\la}$.
Of these,
$H_{\mu\nu}$ is antisymmetric,
$c_{\mu\nu}$, $d_{\mu\nu}$ are traceless,
and all are real.
We note that
the terms involving
$a_\mu$, $b_\mu$,
$e_\mu$, $f_\mu$, $g_{\mu\nu\la}$
break CPT
while those involving
$H_{\mu\nu}$, $c_{\mu\nu}$, $d_{\mu\nu}$
preserve it.
All eight terms are observer Lorentz covariant,
but break particle Lorentz symmetry.
It is possible to eliminate
all the $e_\mu$ and $f_\mu$ terms
and some of the $g_{\mu\nu\la}$ components
to first order
by a field redefinition.
It will suffice here to set
$e_\mu$, $f_\mu$, $g_{\mu\nu\la}$
equal to zero.\ci{ck98}
Superscripts on the couplings will be used where needed to
distinguish,
for example,
electron-positron sector couplings from
proton-antiproton sector couplings.

The CPT- and Lorentz-violating couplings
must all be minuscule
since no breaking of these symmetries
has been observed to date.
A natural suppression scale for these quantities
is the ratio of a light scale
to a scale of order of
the Planck mass
arising in a more fundamental model.

\section{Symmetry Tests with Antihydrogen and Hydrogen}
We have investigated tests of CPT symmetry in \hydrogen\ and \antihydrogen\
using both free and trapped atoms.
For the free case,\ci{bkr99}
the possible signals affecting the 1S-2S transitions
in the context of
the standard-model extension are suppressed
by at least two factors of the fine-structure constant.

\subsection{The 1S to 2S transitions in the trapped case}
Experiments for the creation of
\antihydrogen\ \ci{gab2,holz}
propose to conduct spectroscopic measurements on
\hydrogen\ or \antihydrogen\
held within a magnetic trap
with an axial bias magnetic field.
The Ioffe-Pritchard trap \ci{i-p}
confines states where
the magnetic dipole moment
is opposite in direction to the magnetic field.
The magnetic bias field $B$
splits the
1S and 2S levels
into four
hyperfine Zeeman levels,
which we denote
in order of increasing energy,
by
$\ket{a}_n$, $\ket{b}_n$, $\ket{c}_n$, $\ket{d}_n$,
with principal quantum number $n=1$ or $2$,
for both \hydrogen\ and \ah.

Only the $\ket c$ and $\ket d$
states are trapped,
so it is relevant to consider
transitions involving these states.
For small values of the magnetic field,
transitions between the
$\ket{d}_1$ and $\ket{d}_2$ states are field independent.
It would therefore seem experimentally advantageous
to compare the frequency $\nu^H_d$
for the 1S-2S transition $\ket{d}_1 \rightarrow \ket{d}_2$
in \hydrogen\
with the corresponding frequency $\nu^{\overline{H}}_d$
in \ah,
because the magnetic-field stability
and inhomogeneity
would be less critical.
However,
there are again no unsuppressed frequency shifts
in this transition.
The same holds for \ah.
So, for this transition,
we find
$\de \nu^H_d = \de \nu^{\overline{H}}_d \simeq 0$
at leading order.

Another possibility would be to consider
the 1S-2S transition between
the mixed-spin states $\ket{c}_1$ and $\ket{c}_2$
in \hydrogen\ and \ah.
We find that an unsuppressed frequency shift
occurs in 1S-2S transitions between
the $\ket{c}_1$ and $\ket{c}_2$ states
because the $n$ dependence in the hyperfine splitting produces
a spin-mixing difference between the 1S and 2S levels.
The frequency shift due to
the CPT- and Lorentz-violating terms
is field dependent
with a maximum at about $B \simeq 0.01~T$.
However,
an experimental constraint
is also relevant here.
The 1S-2S transition
$\ket{c}_1 \rightarrow \ket{c}_2$
in \hydrogen\ and \antihydrogen\ is field dependent,
and so the inhomogeneous trapping fields
could lead to significant Zeeman broadening
in the line.

\subsection{Hyperfine transitions in the 1S level}
We next consider
frequency measurements of transitions in the
hyperfine Zeeman sublevels.
The CPT- and Lorentz-violating couplings
in the standard-model extension
give rise to field-dependent energy shifts
of the $\ket a$ and $\ket c$ hyperfine levels
and field-independent shifts
of the $\ket b$ and $\ket d$ hyperfine levels
in the 1S ground state of \h.

A favorable transition is the
$\ket{d}_1 \longrightarrow \ket{c}_1$
line.
At a magnetic field of about 0.65~T,
the frequency shifts in \hydrogen\ and \antihydrogen\
have opposite signs
and are unsuppressed.
The difference in the frequencies
$\nu_{c \rightarrow d}^H$
and
$\nu_{c \rightarrow d}^{\overline{H}}$,
calculated to leading order,
is
$\De \nu_{c \rightarrow d} \equiv
\nu_{c \rightarrow d}^H - \nu_{c \rightarrow d}^{\overline{H}}
\approx - 2 b_3^p / \pi$.
It isolates the CPT-violating coupling $b_3^p$
for the proton
and is therefore of interest as a
clean test of CPT within the standard-model extension.

An appropriate figure of merit
$r^H_{rf,c \rightarrow d}$
can be defined \ci{bkr99}
and it is found that
\beq
r^H_{rf,c \rightarrow d}
\approx
\fr {2\pi |\De \nu_{c \rightarrow d}|} {m_H}
\quad ,
\label{rrf}
\eeq
where $m_H$ is the atomic mass of \h.
Assuming  a frequency resolution of about 1 mHz
would be possible,
we estimate an upper bound of
$
r^H_{rf,c \rightarrow d} \lsim 5 \times 10^{-27}
$.
The corresponding limit on
the CPT- and Lorentz-violating coupling $b_3^p$
would be $|b_3^p| \lsim 10^{-18}$ eV.
This is more than four orders of magnitude better
than bounds attainable from 1S-2S
transitions.

It is worth noting that
the frequency resolution
of high-precision clock-comparison experiments,
which can also constrain Lorentz violation,
lies below 1 $\mu$Hz.
The complex nuclei involved,
however,
make the theoretical interpretation of these experiments
more difficult.\ci{kla}

A feature of the
$\ket{d}_1 \longrightarrow \ket{c}_1$
transition is time-dependence of the shift
due to the changing angle between the $b_j^p$ vector
and the magnetic field of the trap.
Relevant figures of merit for
diurnal-variation signals of this type,
as well as other instantaneous comparisons for
\hydrogen\ and \antihydrogen\
are discussed elsewhere.\ci{bkr99}

\section{Symmetry Tests with Penning Traps}
The behavior of an electron or other charged particle in a Penning trap
is characterized by several frequencies
including the cyclotron and anomaly frequencies,
which can be measured
to precisions better than one part in $10^8$.
For an electron or positron,
the leading-order shifts of these frequencies
in the framework of the standard-model extension are
\bea
\om_c^{e^-} &\approx& \om_c^{e^+} \approx
(1 - c_{00}^e - c_{11}^e - c_{22}^e) \om_c
\quad ,
\label{wcelec} \\
\om_a^{e^\mp} &\approx& \om_a
\mp 2 b_3^e + 2 d_{30}^e m_e + 2 H_{12}^e
\quad .
\label{waelec}
\eea
Here,
$\om_c$ and $\om_a$ denote the
unperturbed frequencies,
while $\om_c^{e^\mp}$ and $\om_a^{e^\mp}$ represent
the frequencies including the corrections.
These two expressions
contain the dominant corrections
in the CPT- and Lorentz-breaking quantities,
and in the magnetic field.

\subsection{Anomalous Magnetic Moments}
Using Eqs.\ \rf{wcelec} and \rf{waelec},
the electron-positron differences
for the cyclotron and anomaly frequencies are
\beq
\De \om_c^e \equiv \om_c^{e^-} - \om_c^{e^+} \approx 0
\quad , \qquad
\De \om_a^e  \equiv \om_a^{e^-} - \om_a^{e^+} \approx - 4 b_3^e
\quad .
\label{delwce}
\eeq
It follows that
the dominant signal for CPT breaking
in Penning-trap $g-2$ experiments is a difference
between the electron and positron anomaly frequencies.
No leading-order contributions appear from
CPT preserving but Lorentz breaking terms.
An appropriate figure of merit can be introduced
in a general context as the ratio of
a CPT-violating electron-positron energy-level difference
and the basic energy scale \ci{bkr97}
\beq
r^e_{\om_a}
\equiv \fr{|{\cal E}_{n,s}^{e^-} - {\cal E}_{n,-s}^{e^+}|}
{{\cal E}_{n,s}^{e^-} }
\quad .
\label{re}
\eeq
Here,
$\cE^{e^-}_{n,s}$ and $\cE^{e^+}_{n,s}$ are energy eigenvalues
for the full Penning-trap hamiltonians,
with quantum numbers $n=0,1,2,\ldots$ and spin $s=\pm 1$.
Within the present nonrelativistic framework
$\cE^{e^-}_{n,s}\to m_e$
and the energy difference
in the numerator becomes half the difference
between the two measured anomaly frequencies,
$\De \om_a^e/2 \approx -2 b_3^e$.
Thus, in the present context
\rf{re} reduces to
\beq
r^e_{\om_a} \approx
\fr{| \De \om_a^e |}{2 m_e}
\approx \fr{|2 b_3^e |}{m_e}
\quad .
\label{reb}
\eeq
If the anomaly frequencies were
measured to an accuracy of 2~Hz,
then this experiment would place a bound of
$r^e_{\om_a} \lsim 10^{-20}$.

It is important to note that
past CPT tests \ci{vd}
have made high-precision comparisons of
the $g$ factors
of electrons and positrons,
to bound the conventional figure of merit
\beq
r_g \equiv \left|\fr{(g_- - g_+)}{g_{\rm av}}\right|
\lsim 2 \times 10^{-12}
\quad .
\label{rg}
\eeq
Within the framework
of the standard-model extension,
however,
CPT is broken without
affecting the electron or positron gyromagnetic ratios.
Thus,
the theoretical value of $r_g$ would be zero
even if CPT were broken,
and $r_g$ used in these experiments is
unsuitable as a CPT figure of merit
in this context.

\subsection{Diurnal anomaly-frequency signals}
Various experiments searching for diurnal signals
are possible.\ci{bkr98}
One option is to search for
a diurnal signal in the anomaly frequency
for an electron alone, or for a positron alone.
In the standard-model extension,
this variation would occur because
the components of the couplings in
Eq.\rf{waelec}
would change as the earth rotates.
Consider
\beq
\De^{e}_{\om_a^{e^-}}
\equiv
\fr {|{\cE}_{0,+1}^{e^-} - {\cE}_{1,- 1}^{e^-}|}
{{\cE}_{0,-1}^{e^-}}
\quad .
\label{Deleorpdnlom}
\eeq
A suitable figure of merit
$r^e_{\om_{a}^-,\rm diurnal}$
may be defined as
the amplitude of the diurnal variation
in $\De^{e}_{\om_a^{e^-}}$.

Data from an experiment confining a single electron
in a Penning trap has recently been analyzed
by Mittleman of the Dehmelt trapping group
at the University of Washington in Seattle, WA.
The bound produced in this experiment is
on a figure of merit related to
$r^e_{\om_{a}^{},\rm diurnal}$.
To search for diurnal variations,
the data was split up into several bins over the sidereal day
according to the orientation
of the experimental magnetic field.
The bound obtained~\ci{rm99}
is
\beq
r^e_{\om_a^-,{\rm diurnal}} \leq 1.6 \times 10^{-21}
\quad .
\eeq
This places a new constraint on a combination
of Lorentz- and CPT-violating quantities in the standard-model
extension.

Another result~\ci{hd99}
has recently been obtained by Dehmelt and collaborators
at the University of Washington.
Using previously obtained data from g-2 experiments
comparing single trapped electrons and single trapped positrons,
a bound of
\beq
r_e < 1.2 \times 10^{-21}
\eeq
was found on a figure of merit closely related to
the one defined in Eq.\ \rf{re}.

\subsection{Hydrogen Ions and Antiprotons}
In Penning-trap experiments comparing
protons and antiprotons,\ci{gg95}
the main measurement uncertainty
is due to the technical difficulties in
precisely reversing the electrostatic potential
on the electrodes when particles of opposite charge are loaded.
To overcome these constraints
Gabrielse and coworkers
refined their experiment to simultaneously trap
an antiproton and a hydrogen ion.\ci{gg99}
The idea is that the H$^-$ ion substitutes for the proton
and corrections are made for the electrons in the ion.
Since both trapped particles have the same charge,
problems of reversing the electric-field polarity
are eliminated.
Measurements of the cyclotron frequency
can be made on each particle independently,
and with greater frequency than previously.
The theoretical value of the difference
$\De \om_c^{H^-}
\equiv\om_c^{H^-} - \om_c^{\bar p}$
is obtained in conventional quantum theory
using established precision measurements of the electron mass
and the H$^-$ binding energy.
Comparison of this theoretical value
with the experimental result for
$\De \om_c^{H^-}$
Provides
a proton-antiproton charge-to-mass test.

Employing this technique,
Gabrielse's group recently
obtained the result
\beq
r_{q/m}^p \equiv
\left|\fr
{\left[ (q_p/m_p)- (q_{\overline{p}}/m_{\overline{p}})\right]}
{(q/m)_{\rm av}}
\right|
\lsim
9 \times 10^{-11}
\label{rqmpNew}
\eeq
for the proton-antiproton charge-to-mass ratio
comparison.\ci{gg99}
Apart from this ten-fold improvement
over the previous value,
the experiment also placed the bound
\beq
r^{H^-}_{\om_c} \lsim 4 \times 10^{-26}
\label{Hminbound}
\eeq
on a Lorentz-violation
figure of merit \ci{bkr98}
within the framework of the standard-model extension.

\section{Discussion}
Other possibilities for detecting signals of CPT and Lorentz
violation not discussed here
include diurnal frequency variations and
cyclotron-frequency measurements in Penning traps,\ci{bkr98}
and alternative transitions in \hydrogen\ and \antihydrogen\ atoms.
Furthermore,
the electron-positron Penning-trap experiments
could in principle be performed with protons and antiprotons.
Experiments involving boosted particles,
for example \hydrogen\ or \antihydrogen\ beams,\ci{blanPRD}
have an enhanced sensitivity to timelike components
\ci{ak98}
like $b_0^e$ and $b_0^p$.
In this summary,
several interesting cases of signals that might occur
in Penning-trap and \hydrogen\ or \antihydrogen\ experiments
have been discussed.
The theoretical investigation is based on the framework provided
by the standard-model extension.\ci{ck98}
Three experiments mentioned above have ruled out signals
of Lorentz and CPT violation
at current precisions,
but future investigations could still find evidence of
the breaking of these fundamental symmetries
either through improved precisions
or through as yet unperformed experiments.

\section*{Acknowledgments}
This work was partially supported by the United States Department of Energy
under grant DE-FG02-91ER40661.
I thank the NSF for
funding to attend this meeting.

\end{document}